# Dual-Port Dynamically Reconfigurable Battery with Semi-Controlled and Fully-Controlled Outputs

N. Tashakor, *Student Member, IEEE*, J. Kacetl, *Student Member, IEEE*, J. Fang, *Member, IEEE*,
Z. Li, *Member, IEEE* and S. Goetz, *Member, IEEE*

*Abstract*—Modular multilevel converters (MMC) and cascaded H-bridge (CHB) converters are an established concept in ultra-high voltage systems. In combination with batteries, these circuits allow dynamically changing the series/parallel configuration of subportions of the battery as so-called modular battery-integrated converters or reconfigurable batteries, and are being discussed for grid-storage and electromobility applications. A large body of research focuses on such circuits for supplying a single load, such as a motor for electric drives. Modularity, failure tolerance, less dependence on the weakest element of a battery pack, higher controllability, and better efficiency are the main incentives behind this pursuit. However, most studies neglect the auxiliary loads which require isolation from the high-voltage battery. This paper proposes a simple topology and controller that can fork off a second (galvanically isolated) output of a reconfigurable dc battery. The proposed system provides a nonisolated semi-controlled port for the dc link to maintain the operating point of the main inverter(s) close to optimal, while fully controlling an isolated output for the auxiliaries per the safety regulations. The proposed system does not require additional active switches for the auxiliary port and can operate with a wide range of voltages. Simulation and experiments verify the developed analysis.

## I. INTRODUCTION

ALTHOUGH recent technological developments as well as environmental incentives have sped up the electric vehicle's expansion into the market, many of the traditional challenges remain. In the recent years, a combination of technological progress and request for larger ranges by drivers have almost tripled the capacity of the battery packs used in modern electric vehicles [1]. Today, an EV is powered by a mixed serial and parallel connection of literally hundreds of cells. In addition to the increased capacity, a trend toward higher voltage levels is observed that leads to increasing the share of serial connection in batteries with the same energy capacity [2]. Jung discusses the advantages of a higher voltage battery pack (i.e., 800 V) including lower weight, better efficiency, and faster charging [3]. However, higher numbers of serial connections also introduce problems including more complex monitoring, the need for several voltage levels for legacy systems among loads and charging infrastructure or due to safety, lower efficiency of the inverters at partial load, protection, and increasing chances of poor battery cells in the pack due to manufacturing tolerances, which limit the overall performance [4-6].

Figure 1(a) depicts the electrical circuit of a conventional EV. The high-power drive system includes high-voltage batteries and often a dc/dc converter that supplies the dc link of the main inverter(s) [7]. Additionally, a second low-voltage isolated output supplies the auxiliaries (e.g., lights and HVAC system). As many studies show, it is possible to further shift the operating point of the inverters by actively regulating the dc-link voltage [8-10]. Regulating the dc-link voltage improves the efficiency of the system, but the other problems including the balancing battery cells as well as fault tolerance persist [11, 12].

The recent advancement in the field of modular systems and the performance gains of low-voltage transistors have stimulated the development of concepts for dynamically reconfigurable battery systems, which are sometimes also called modular battery-integrated converters or split battery systems [13-20]. These systems break the previously hard-wired battery pack, which was limited by its weakest element, into subunits, often below 100 V, rarely on the cell level, and add electronic or rarer electromechanical switches to enable a circuit reconfiguration [21, 22]. Simplified derivates of such concepts are under series development in the vehicle industry [23-25]. A reconfigurable battery can provide balancing functionality, better fault tolerance, and faster output regulation [26-29]. Furthermore, the overall efficiency can be improved [30, 31]. However, while many different topologies have been proposed for driving the motors, most of the modular topologies ignore auxiliary load requirements of an EV and therefore need a separate set of battery and converter. Due to current safety requirements, auxiliary supplies below the safe extra-low voltage level (SELV) must be isolated from the high voltage [32, 33].

Gan et al. propose a battery-integrated MMC for the battery system to provide a fault-tolerant, highly modular, and fully controllable dc-link voltage for driving a switched reluctance motor (SRM). However, this topology does not consider the auxiliaries, and completely independent auxiliary power modules are necessary [13]. In order to solve the problem of galvanic isolation, Kandasamy et al. propose an inductively coupled battery-integrated full-bridge inverter [34]. However, multiple high-frequency transformers as well as a high number of active components reduce the overall efficiency and increase the cost as well as size of the system.

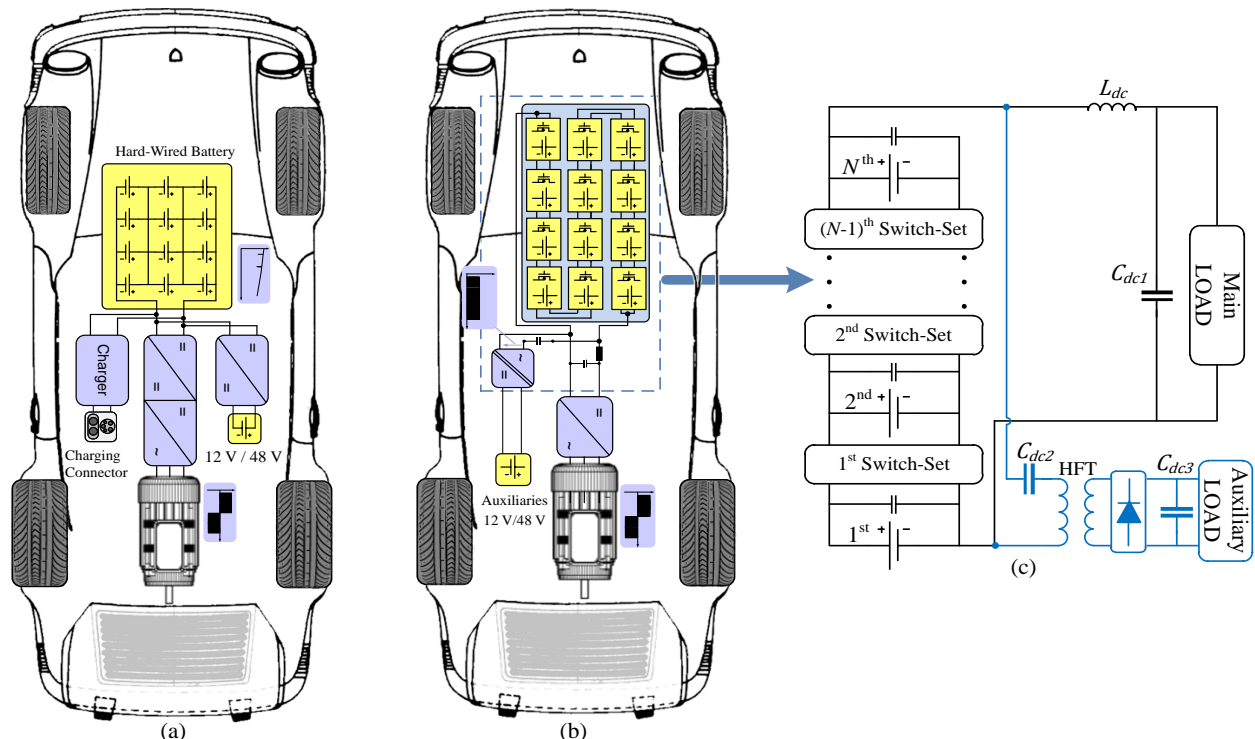

Fig. 1. (a) electrical circuit of a conventional electric drive; (b) electrical circuit of the proposed system; (c) macro-level structure of the proposed multi-port system

Although multi-port systems are not a new concept in grid-storage applications, most of the available topologies are not fit for reconfigurable battery systems in EVs [35, 36]. This paper fills this gap by presenting a dual-port dynamically reconfigurable battery. The proposed topology makes use of the already available half-bridge modules to provide a semi-controlled dc-link voltage for the main drive system and concurrently generate an isolated fully-controlled dc voltage for the auxiliaries. The semi-controlled dc-link system can shift the operating point of the motor inverters close to optimal, while the fully-controlled isolated output can provide a constant voltage and/or controlled current as expected by auxiliary loads under varying currents. Additionally, no extra switches are required to generate the controlled isolated voltage, which further adds to the appeal of this topology. Furthermore, the proposed system does not introduce any extra imbalance and loads all the modules equally.

The following presents the topology and an analysis for the dual-port system in Section II. Section III describes the control strategy for the main as well as auxiliary ports, and Section IV verifies them through simulation and experiments. Finally, Section V concludes the paper with remarks.

## II. ANALYSIS OF THE PROPOSED DUAL-PORT RECONFIGURABLE BATTERY

Reconfigurable batteries are not a new concept but can refer to many different macro- and micro-topologies available in the literature [11, 12, 37-39]. Various string connections can provide dc, single-phase, and multi-phase structures with specific features [14, 15]. Figure 1(b) shows the macro-structure of the system. The modular topology in combination with the inductor ($L$) and the dc capacitor ($C_{dc1}$) forms a dc–dc buck converter (i.e., circuit in black) that can control the dc-link voltage of the traction inverters. Additionally, the dc capacitors ($C_{dc2}$ and $C_{dc3}$), the high-frequency transformer, and the diode-bridge form a second isolated output port for the auxiliary loads of the EV (see the circuit in blue).

Different topologies can serve as electronic basis for battery modules, where half-bridge has the simplest form that can provide a multilevel output voltage using low-voltage switches [39-41]. However, other topologies such as the three-switch and dual HB topologies provide additional parallel connectivity across modules [20, 42-44]. Figure 2 shows the two simplest possible module topologies that are applicable with Fig. 1(c) as well as their different modes of operation. Although parallel and bypass connection result in almost identical outputs in Fig. 3, the bypass mode is not recommended except during fault or highly unbalanced conditions, since parallel connection results in better efficiency [45].

### A. High-Power Modular DC/DC Converter

The modular buck converter is responsible for maintaining the dc-link voltage of the traction inverters within the optimal range, based on the operation point of the traction system and its corresponding efficiency map [46, 47]. Since determining the optimum operating region of the system is not the main contribution of this work, we consider the desired dc-link voltage as reference value for our system.

The well-known phase shifted carrier (PSC) modulation generates the switching signals for the modules in the string [48, 49]. Each switch-set in Fig. 1(c) corresponds to one carrier. Therefore, for $N$ battery modules in Fig. 2, $(N-1)$ carriers are required. The carrier waveforms are compared with one universal modulation index to generate the switching pulses for its respective switch set. Fig. 3 shows an intuitive representation of a system with $(N-1)$ carriers.

Therefore, the dc-link voltage of the main load is
$$V_{dc1} = (1 + m(N_{arm} - 1))V_m, \quad (1)$$
where $V_m$ is the voltage of one module.

### B. Auxiliary Power Unit

This section derives the necessary equations to estimate the output of the isolated auxiliary port with respect to the modulation index $m$. Fig. 4(a) shows the circuit of this port. Depending on the value of $m$ and $N$, the output of the system with PSC modulation is similar to the summation of one constant voltage ($V_{base}$) and a PWM controlled voltage ($V_{pulse} \in [V_m, 0]$) with the effective switching rate of $(N-1)f_{sw}$, where $f_{sw}$ is the frequency of one carrier. The value of the base voltage can be calculated using
$$V_{base} = [\text{floor}(m(N_{arm} - 1) + 1)]V_m. \quad (2)$$

Figure 4(b) shows the effective carrier waveform for the PWM voltage in an intuitive manner. The resulted duty cycle of the pulses follows
$$D = m(N_{arm} - 1) - \text{floor}(m(N_{arm} - 1)), \quad (3)$$
which indicates a repeating sequence with respect to $m$.

At each operating point, capacitor $C_{dc2}$ is charged up to $V_{dc1}$ and the waveform of input voltage ($V_{in}$) to the switching transformer includes a positive pulse ($V_{p+}$) and a negative pulse ($V_{p-}$), which are calculated per
$$V_{p+} = (1 - D)(V_m - \Delta V_r), \quad (4)$$
$$V_{p-} = -D(V_m - \Delta V_r), \quad (5)$$
as shown in Fig. 5 intuitively, where $\Delta V_r$ is the voltage ripple on capacitor . Varying $m$ changes the value of $D$ which in turn affects the shape of the input voltage of the transformer. Two situations are possible depending on $D$ as follows
$$\begin{cases} D \le 0.5 \\ D > 0.5 \end{cases} \Rightarrow \begin{matrix} V_{p+} \ge V_{p-} \\ V_{p+} < V_{p-} \end{matrix}. \quad (6)$$

For $D < 0.5$, the positive pulse is larger. The capacitor is charged when $0 \le t \le DT_{sw,eff}$. With moving all the components to the secondary side of the switching transformer, the relation between $i_1$, $i_2$, and load current $I_{load}$ follows
$$i'_1 = \frac{N_2}{N_1} \cdot i_2, \quad (7)$$
$$i_2 = \frac{I_{load}}{D}, \quad (8)$$
where $N_1$ and $N_2$ are respectively winding coefficients of the primary and secondary sides.

Figure 6(a) shows the equivalent circuit of the APU when $D \le 0.5$. Applying KVL results in
$$\left(V_{p+} - \frac{N_2}{N_1} \cdot \frac{I_{load}}{D} \cdot r_{eq1}\right) \cdot \frac{N_2}{N_1} - r_{eq2} \cdot \frac{I_{load}}{D} - 2V_{fd} = V_o, \quad (9)$$
where $r_{eq1} = r_{C_{dc2}} + r_{L1}$ and $r_{eq2} = r_{L2} + 2r_{fd}$.

When $DT_{sw,eff} \le t \le T_{sw,eff}$, the diode-bridge is open circuit and the $C_{dc3}$ discharges to supply the load following
$$V_{p-} - r_{eq1} \cdot i_1 - R_C i_1 = 0. \quad (10)$$

From these two states, the steady-state output voltage can be calculated based on the value of $V_{p+}$. Substituting (4) in (9) results in
$$\left((1-D)(V_m - \Delta V_r) - \frac{N_2}{N_1}\frac{I_{load}}{D}r_{eq1}\right)\frac{N_2}{N_1} - r_{eq2}\frac{I_{load}}{D} - 2V_{fd} = V_{dc2}, \quad (11)$$
where $I_{load} = \frac{V_{load}}{R_{load}}$ and $V_m$ is the voltage of an individual

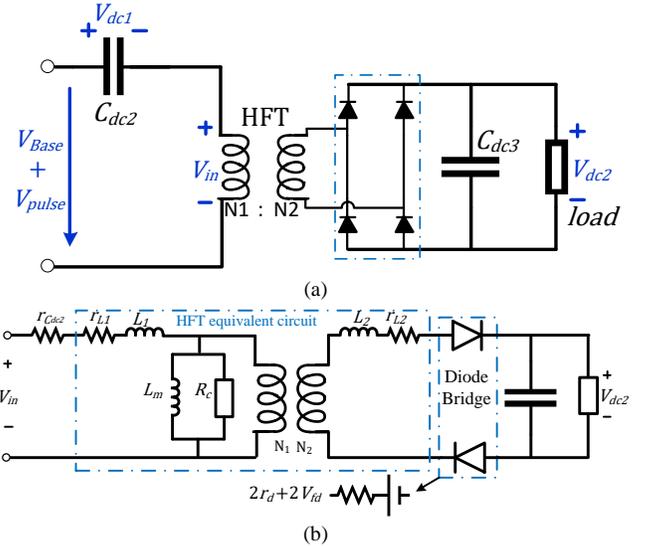

Fig. 4. (a) Circuit diagram of the auxiliary unit; (b) equivalent electrical circuit with a positive $V_{in}$

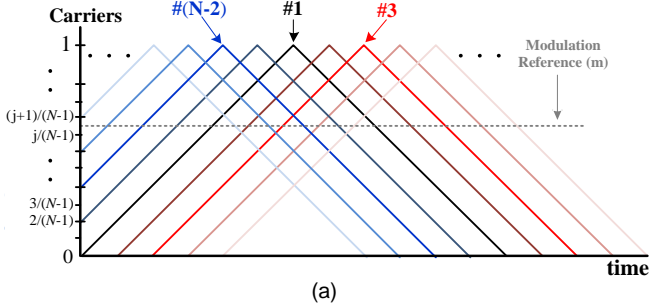

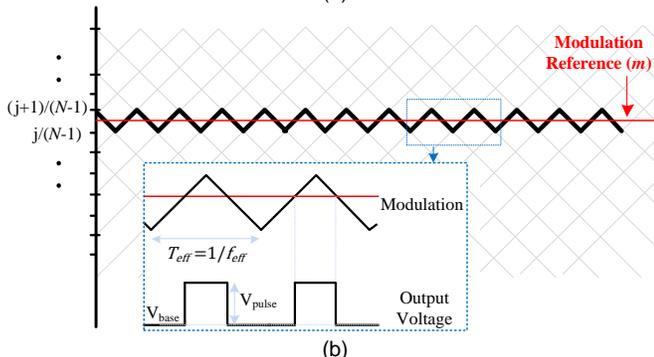

Fig. 3. Intuitive representation of PSC modulation: (a) PSC carriers; (b) equivalent carriers in the case of symmetrical PSC

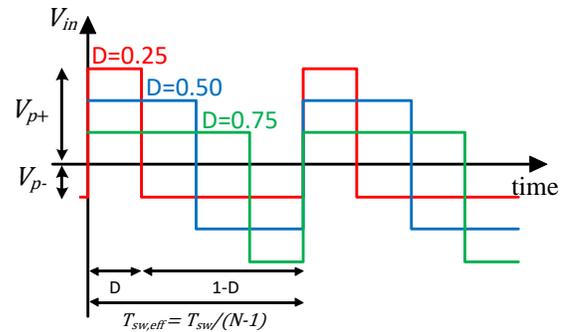

Fig. 5. intuitive representation of $V_{in}$ with different $D$

battery module. Through some simplification and manipulation, (11) is rewritten as

$$\frac{N_2}{N_1}(1-D)(V_m - \Delta V_r) - 2V_{fd} = V_{dc2}(R_{eq} + 1), \quad (12)$$

where $R_{eq}$ is the total equivalent resistance of the system and is calculated per

$$R_{eq} = \left(\left(\frac{N_2}{N_1}\right)^2 r_{eq1} + r_{eq2}\right)\frac{1}{D \cdot R_{load}}. \quad (13)$$

Finally, the output gain of the system for $D \leq 0.5$ is

$$\frac{V_{dc2}}{V_m} = \frac{(1-D)\frac{N_2}{N_1}(1-\frac{\Delta V_r}{V_m}) - \frac{2V_{fd}}{V_m}}{R_{eq}+1}, \quad (14)$$

which can be further simplified by neglecting the $V_{fd}$ and $\Delta V_r$ into

$$\frac{V_{dc2}}{V_m} = \left(\frac{1-D}{R_{eq}+1}\right)\frac{N_2}{N_1}. \quad (15)$$

For $D > 0.5$, the negative pulse is larger. Therefore, the diode bridge is open circuit during positive pulses ($0 \leq t \leq DT_{sw,eff}$) and charges during the negative pulses ($DT_{sw,eff} \leq t \leq T_{sw,eff}$). During this condition, the relation between the secondary current and the load current follows

$$i_2 = \frac{I_{load}}{1-D}. \quad (16)$$

Figure 6(b) presents the electrical equivalent circuits of the system during this condition and similarly KVL results in

$$\left(V_{p-} - \frac{N_2}{N_1}\left(\frac{I_{load}}{1-D}\right)r_{eq1}\right)\left(\frac{N_2}{N_1}\right) - r_{eq2}\frac{I_{load}}{1-D} - 2V_{fd} = V_{dc2}, \quad (17)$$

which can be simplified to

$$\frac{V_{dc2}}{V_m} = \frac{D\frac{N_2}{N_1}(1-\frac{\Delta V_r}{V_m}) - \frac{2V_{fd}}{V_m}}{R_{eq}+1}, \quad (18)$$

with $R_{eq} = \left(\left(\frac{N_2}{N_1}\right)^2 r_{eq1} + r_{eq2}\right)\frac{1}{(1-D)R_{load}}$.

Not considering $V_{fd}$ and $\Delta V_r$ simplifies (18) further into

$$\frac{V_{dc2}}{V_m} = \left(\frac{D}{R_{eq}+1}\right)\frac{N_2}{N_1}. \quad (19)$$

Based on (14) and (19), $\forall D \in [0,1] \mid V_{dc2} \in [0.5 V_m, V_m]\frac{N_2}{N_1}$, and for each $V_{dc2}$, there are two possible $D$ values. Furthermore, based on (3), each value of $D$ entails $(N-1)$ possible values for $m$. Therefore, at each operating point, there are $2(N-1)$ values for $m$ which lead to an identical voltage in the output of the auxiliary power unit ($V_{dc2}$), while resulting into completely different output in the dc-link voltage ($V_{dc1}$) of the inverters. The extra degree of freedom can be exploited to maintain both outputs within the desired range, which we will discuss in the next section.

## III. PROPOSED CONTROLLER AND SYSTEM DESIGN

As stated above, the two main objectives of the system are to maintain the dc-link voltage of the traction system ($V_{dc1}$) within the optimum operating range of the motor–inverter set, while actively controlling the output voltage of the auxiliary power unit at its reference value. The optimum operating dc-link voltage for inverter ($V_{dc1}^{ref}$) is assumed as a reference from the higher-level control loops, and the reference output voltage of auxiliary power unit ($V_{dc2}$) is considered constant (e.g., in most EVs either around 12 V per LV 124 standard or 48 V per LV 148 and VDA 320 standards).

The values of module voltage and the expected voltage of the auxiliary power unit determine the transformer ratio $\left(\frac{N_2}{N_1}\right)$.

Based on (13) and (18) the suitable range of $\frac{N_2}{N_1}$ is

$$\frac{(R_{eq}+1)(V_{dc2}-V_{fd})}{0.95(V_m - \Delta V_r)} \leq \frac{N_2}{N_1} \leq \frac{(R_{eq}+1)(V_{dc2}-V_{fd})}{0.5(V_m - \Delta V_r)}. \quad (20)$$

Although in theory (20) shows a large range to calculate the transformer ratio since the voltage of the battery modules can vary according to their state of health (e.g., for many Li-Ion cell chemistries $0.85 V_{rated} \leq V_m \leq 1.2 V_{rated}$, such as LiPo), a more practical relation is

$$\frac{(R_{eq}+1)(V_{dc2}-V_{fd})}{0.95(V_{m,min} - \Delta V_r)} \leq \frac{N_2}{N_1} \leq \frac{(R_{eq}+1)(V_{dc2}-V_{fd})}{0.5(V_{m,max} - \Delta V_r)}, \quad (21)$$

where $V_{m,min}$ and $V_{m,max}$ are the minimum and maximum operating voltage of each module.

Furthermore, due to $R_{eq}$ in (19) and (14), the gain of the system is not completely linear and as $D$ goes to the upper and lower bounds (i.e., close to one and zero), the effect of parasitic resistances on the system increases linearly. As an example, Fig. 7 shows the ideal as well as non-ideal calculated gains of the simulated system in the next section and compares them with actual simulation results.

Based on Fig. 7, designing the normal operating point of the system close to $D = 0.5$ improves the controllable range of the system, and $D < 0.2$ or $D > 0.8$ deteriorates the system performance. Therefore, the upper boundary of $\frac{N_2}{N_1}$ in (21) is the optimal value. Replacing $R_{eq}$ with (13) and solving for $\frac{N_2}{N_1}$ gives the suitable value for the transformer ratio.

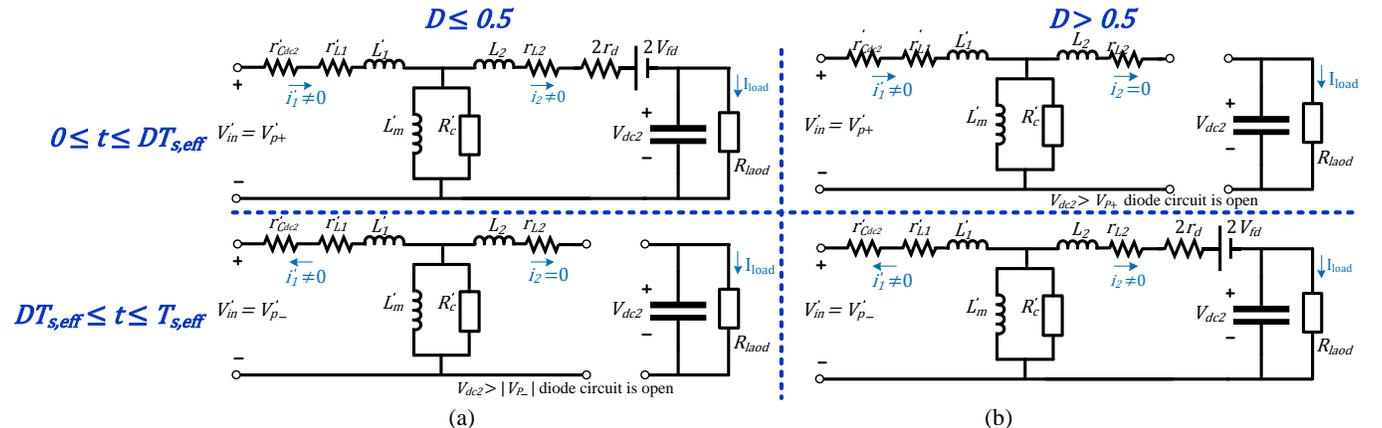

Fig. 6. The equivalent circuit of the system: (a) $D \geq 0.5$, (b) $D < 0.5$

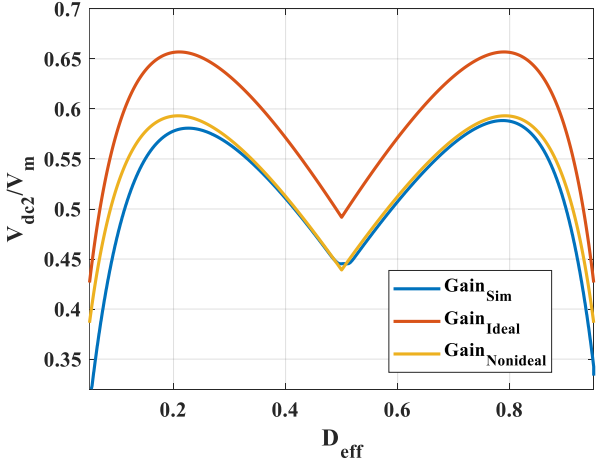

Fig. 7. Gain variations of the auxiliary power unit with respect to $D$

Capacitor $C_{dc2}$ is responsible for decoupling the dc and switching components of the voltage. As seen in (14) and (18), the voltage ripple of capacitor $C_{dc2}$ can significantly affect the system behavior. Using an analysis similar to the design of transformer ratio, the minimum capacitor for the system is

$$C_{dc2} = \frac{P_{max2}}{(N-1)V_{dc2}\Delta V_r f_{sw}} \frac{N_2}{N_1}, \quad (22)$$

where $P_{max2}$ is the maximum output power of the auxiliary unit. Similarly, the minimum capacitance of capacitor $C_{dc3}$ is

$$C_{dc3} = \frac{P_{max2}}{(N-1)V_2 \Delta V_r f_{sw}}. \quad (23)$$

Figure 8 provides the proposed control algorithm for the dual port system. At each instance, the reference values for $V_{dc1}$ is provided by the efficiency maps of the system. The value of $V_{dc2}^{ref}$ is the rated voltage of the auxiliary power unit. $V_{dc2}$ is the measured voltage at the output of the auxiliary power unit and $\overline{V_m}$ is the average operating voltage of the modules. The output voltage of the auxiliary unit for $D$ and $(1-D)$ is identical. Therefore, the algorithm determines the modulation index ($m$) according to (3) that its resulted $D$ is equal to $D^*$ or $(1-D^*)$ and minimizes the difference between $V_{dc1}$ and $V_{dc1}^{ref}$. Hence, the output voltage of the auxiliary unit is fully controlled, and the dc-link voltage of the inverters maintained within a small boundary of the optimal point.

Based on the number of the modules and the transformer ratio, the maximum deviation of the dc-link voltage from $V_{dc1}^{ref}$ is

$$\Delta V_{max} \leq \frac{0.5}{1-N} V_m. \quad (24)$$

Additionally, a hysteresis block in the input of the PI controller reduces fluctuations.

## IV. RESULTS AND DISCUSSION

### A. Simulation Results

MATLAB/Simulink serves to simulate a system with ten modules, where Table I provides its main parameters. In the simulation, batteries are modelled using a simplified electrical equivalent circuit consisting of an internal resistance as well as a constant dc voltage source. The modular battery is feeding a variable load with a variable reference voltage. The rated voltage of the battery is 91 V, which can fluctuate between 80 V to 105 V. The 48 V supplies are becoming established in many executive and sports cars, so we set $V_{dc2}^{ref} = 48$ V. Based on the module voltage range as well as $V_{dc2}^{ref}$ value, (22) determines the ratio of the transformer to approximately 1.13.

The system is simulated for rated battery voltages. Figure 9 shows the voltage and current waveforms for the first and second outputs. During the simulation the optimal reference voltage and also demand power for the first output as well as the demand power of the second output are varied. The controller can provide a fixed $V_{dc2}$ under stark variations of the operation point of both outputs. Additionally, the voltage of the first output closely follows the optimal value provided as a reference to the controller. The steady-state ripple of $V_{dc1}$ and $V_{dc2}$ is below 3 % and 1 %, respectively.

Figure 10 shows the modulation signals as well as the voltage deviation of both outputs from their respective goals. Although there are some transients at the beginning of each step, the steady-state error for the second output is less than 0.2 %. Similarly, maximum voltage deviation of $V_{dc1}$ from its optimal value is below 6 %.

### B. Experimental Results

We built a prototype reconfigurable battery pack with five modules. An FPGA-based rapid prototyping controller (sbRIO 9726) implements the proposed control algorithms as well as the PSC modulation and the measurements are recorded using an eight-channel oscilloscope from LeCroy. An isolated transducer (LV25P due to its low settling time) with an analog amplifier provides the isolated feed-back of the second dc output. Each battery module consists of six series cells that provide a 24 V open-circuit voltage. An RL load with the resistance of 5.5 Ω is connected to the first dc output, while an electronic load is connected to the second isolated output. Table II summarizes the parameters of the laboratory setup and Fig. 11 shows the laboratory testbench.

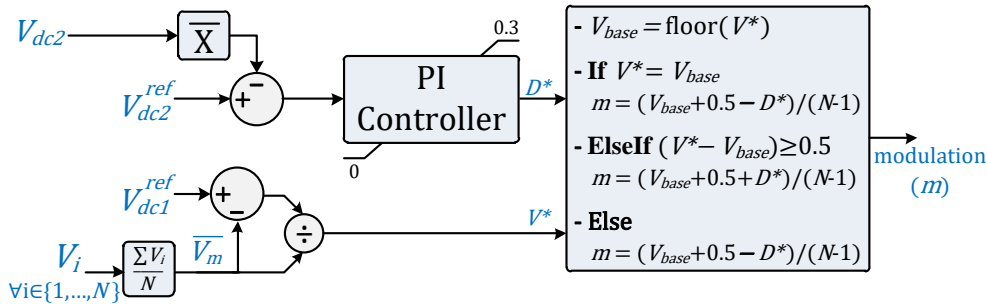

Fig. 8. The proposed control algorithm for the dual-port system

TABLE I
PARAMETERS OF THE SIMULATED SYSTEM

| PARAMETER | VALUE |
|---|---|
| $V_{dc1}$ | 400 – 800 [V] |
| $C_{dc1}$ | 20 [μF] |
| $L_{dc}$ | 23 [μH] |
| $C_{dc2}$ | 348 [μF] |
| $C_{dc3}$ | 5.4 [mF] |
| $R_{ldc}$ | 10 [mΩ] |
| $P_{load,1}$ | 300 [kW] |
| $P_{load,2}$ | 5 [kW] |
| $R_{ds}, R_d$ | 1 [mΩ] |
| $V_m$ | 82 – 103 [V] |
| $r_{bt,1\sim 8}$ | 5 [mΩ] |

The desired voltage of the first dc output ($V_{dc1,ref}$) is provided as an input to the FPGA controller, while the reference output of the auxiliary port can be only 0 V when it is turned off or 12 V when operating. The controller uses the proposed algorithm to determine the most suitable modulation index and then generates the switch signals for all modules. The outputs are recorded by the oscilloscope and later plotted in MATLAB. Figure 12 shows the measurements for output voltages and currents of the main output port. Figure 13 shows the voltage and current at the terminal of the auxiliaries. While the output voltage of the main port is always close to its desired value, it does not fully converge. On the other hand, neglecting minor transients, the output voltage of the auxiliaries fully follows its reference value.

Figure 14 shows the deviations of the output voltages from the reference values. The output voltage of the isolated output ($V_{dc2}$) stays below the 2 % mark. Concurrently, even with relatively large deviations in the reference voltage of the nonisolated output, the provided output voltage is within a 6 % boundary, which is significant improvements compared to the conventional systems with fixed dc-link voltage.

## V. CONCLUSION

This paper proposes a multi-port reconfigurable battery for e-mobility application that can fork off a second (galvanically isolated) dc voltage off a reconfigurable dc battery and control power in both using the extra degrees of freedom offered by the reconfigurable battery. The generated nonisolated semi-controlled output voltage supplies the traction system of the electric vehicle, while a fully-controlled lower power isolated output supplies the auxiliary system. For the auxiliaries, no extra active or controlled components are necessary, which further adds to the appeal of this technique. The nonisolated traction output is semi-controlled and still stays within 7 % of the rated voltage in contrast to >40 % in conventional hard-wired battery packs to enable a tightly controlled second, isolated output despite limited degrees of freedom.

The paper analyzes the behavior of the system and provides design guidelines for both controller and topology. The pro-vided analysis as well as simulation and real measurements support the applicability and performance of the proposed topology and method. The output of the auxiliary port can be maintained within a 2 % boundary of its reference voltage, while the semi-controlled output voltage is maintained within a 7 % boundary. The accuracy of the semi-controlled output can be further improved as the number of existing modules increases.

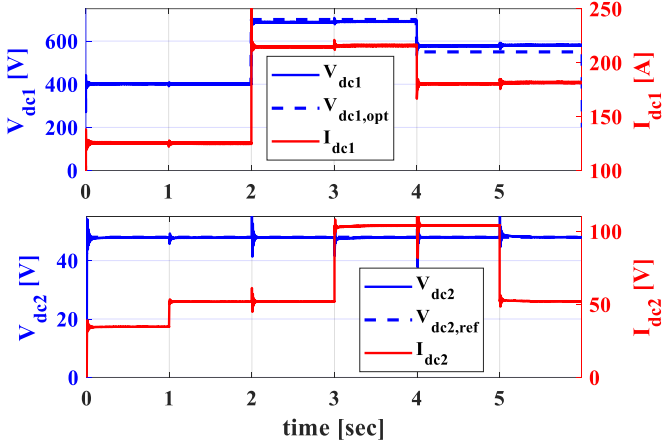

Fig. 9. Voltage and current waveforms for the simulated system

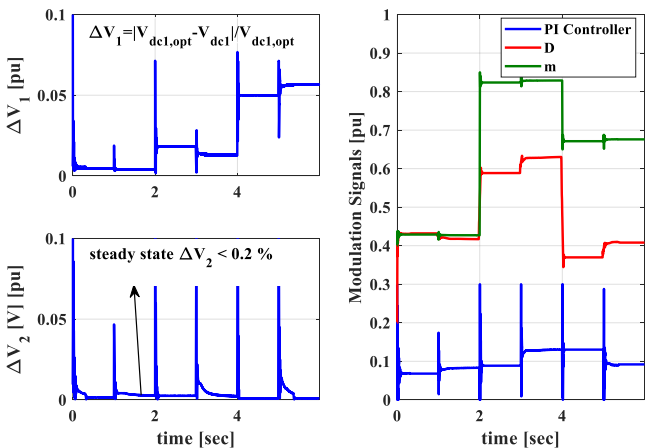

Fig. 10. The control signals as well as the outputs' deviations from reference values

TABLE II
PARAMETERS OF THE LABORATORY SETUP

| PARAMETER | VALUE |
|---|---|
| $V_{dc1}$ | 20 – 100 [V] |
| $C_{dc1}$ | 100 [μF] |
| $L_{dc}$ | 200 [μH] |
| $C_{dc2}$ | 940 [μF] |
| $C_{dc3}$ | 2.7 [mF] |
| $R_{ldc}$ | 50 [mΩ] |
| $P_{load,1}$ | 2 [kW] |
| $P_{load,2}$ | 36 [W] |
| $R_{ds}, R_d$ | 1 [mΩ] |
| $V_m$ | 22 – 25.2 [V] |
| $r_{bt,1\sim 8}$ | 20 [mΩ] |

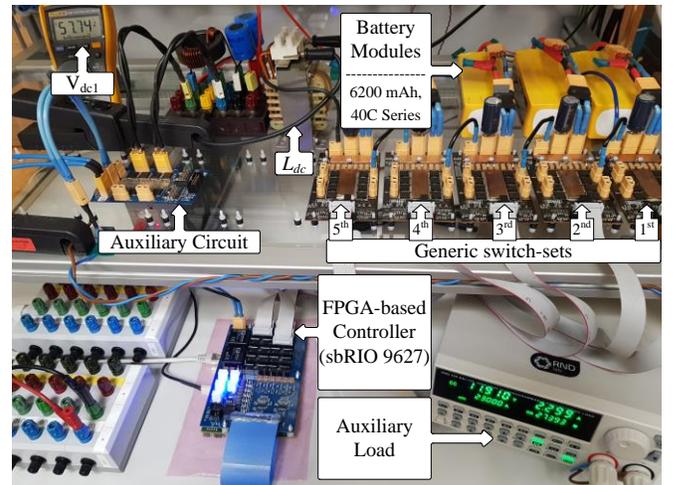

Fig. 11. Laboratory testbench

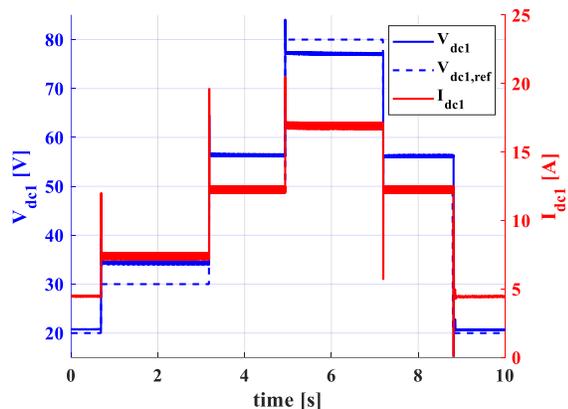

Fig. 12. The measured voltage and currents at the semi-controlled main port

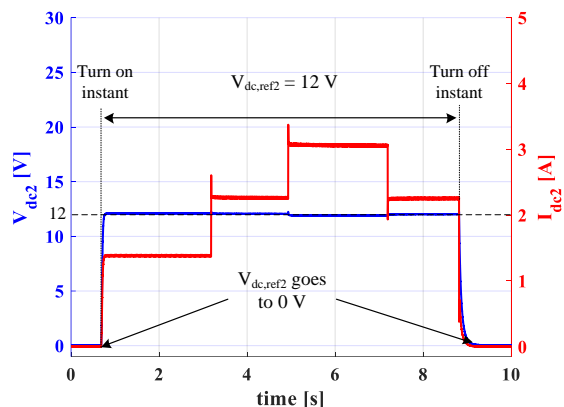

Fig. 13. The measured voltage and currents at the auxiliaries